\documentclass[conference]{IEEEtran}
\IEEEoverridecommandlockouts
\usepackage{amsmath,amsfonts}
\usepackage{amssymb}
\usepackage{array}
\usepackage[caption=false,font=normalsize,labelfont=sf,textfont=sf]{subfig}
\usepackage{textcomp}
\usepackage{stfloats}
\usepackage{url}
\usepackage{graphicx}
\usepackage{bm}
\usepackage{amsthm}
\usepackage{mathrsfs}

\newcommand{\diag}{\mr{diag}}

\newcommand{\mr}{\mathrm}

\newcommand{\BE}{\begin{equation}}
\newcommand{\EE}{\end{equation}}
\newcommand{\BS}{\begin{subequations}}
\newcommand{\ES}{\end{subequations}}
\renewcommand{\bf}{\bm}

\newtheorem{definition}{Definition}
\newtheorem{lemma}{Lemma}

\begin{document}

\title{A Universal Random Precoding Framework for MIMO Systems}

\author{
\IEEEauthorblockN{
Jiazhen Dong$^{1,2}$, Lei Liu$^{1,2}$, Xiaojun Yuan$^{3}$, Baoming Bai$^{2}$
}
\IEEEauthorblockA{
$^{1}$College of Information Science and Electronic Engineering, Zhejiang University, China\\
$^{2}$State Key Laboratory of Integrated Services Networks, Xidian University, China\\
$^{3}$National Key Laboratory of Wireless Communications,\\
University of Electronic Science and Technology of China, China\\
Email: \{dongjiazhen, lei\_liu\}@zju.edu.cn,
xjyuan@uestc.edu.cn,
bmbai@mail.xidian.edu.cn
}
\thanks{
The work of J. Dong and L. Liu was supported in part by the National
Natural Science Foundation of China (NSFC) under Grants 62394292 and
62301485, in part by the Zhejiang Provincial Natural Science Foundation
under Grant LZ25F010002, and in part by the State Key Laboratory of
Integrated Services Networks under Grant ISN25-10.
The work of B. Bai was supported in part by the NSFC under Grant U254120087.

\par\medskip
\copyright{} 2026 IEEE. Personal use of this material is permitted.
Permission from IEEE must be obtained for all other uses, in any current
or future media, including reprinting/republishing this material for
advertising or promotional purposes, creating new collective works, for
resale or redistribution to servers or lists, or reuse of any copyrighted
component of this work in other works.
}
}

\maketitle

\begin{abstract}
Current wireless systems combat inter-symbol interference (ISI) by diagonalizing or sparsifying the channel matrix, yet they remain vulnerable to selective fading. To address this, we propose a universal random precoding (RP) transmission framework based on the universality class. RP leverages random transforms to statistically exploit all subchannels and construct an equivalent channel belonging to the universality class, thereby enhancing diversity gain while maintaining backward compatibility with existing waveforms. Low-complexity implementations include the randomly permuted fast transform (FT-RP) and the interleaved block-sparse fast transform (IBSFT-RP). A cross-domain OAMP/MAMP (CD-OAMP/MAMP) detector is designed for RP systems, which is replica maximum \textit{a posteriori} (MAP)-optimal according to state evolution (SE). Simulation results on MIMO systems demonstrate that RP with CD-OAMP/MAMP achieves near-RM performance with much lower complexity, with additional benefits of flexible compression ratios for spectral efficiency.
\end{abstract}


\section{Introduction}

Channel fading is a major challenge in wireless communication. Most current waveforms sparsify the equivalent channel to simplify detector design. OFDM, used in 4G/5G, diagonalizes the channel in the frequency domain so each subcarrier can be detected independently. Orthogonal time-frequency space (OTFS) \cite{hadani2017orthogonal,wei2021orthogonal} modulation and affine frequency division multiplexing (AFDM) \cite{9562168} modulation sparsify the equivalent channel in the delay-Doppler or warped time-frequency domain, respectively. While AFDM achieves full diversity under exponential-complexity Maximum Likelihood (ML) detection, low-complexity detection remains challenging. Several detectors such as CD-OAMP \cite{9536449} and DD-OAMP \cite{10043628} have been proposed for OTFS and AFDM, but none can be proved maximum \textit{a posteriori} (MAP)-optimal, since OTFS and AFDM generally fail to make the equivalent channel matrix right-unitarily invariant \cite{liu2024capacity,liu2023oamp}.

Random multiplexing (RM) \cite{Liu2025RandomMultiplexing} overcomes these limitations by applying a random modulation matrix, making the equivalent channel statistically smooth. This achieves substantial diversity gain and enables the replica MAP-optimal CD-MAMP detector \cite{10522098,liu2022memory}. IFDM \cite{10522098} is a special case of RM. 
Furthermore, performance of such iterative detectors can be also enhanced using random transforms like the Energy Spreading Transform (EST) \cite{1561586,6857421} and Hadamard-Haar Transform (HHT) \cite{9037251}. These transforms improve the Gaussian properties of the signal, which is known to facilitate more reliable message passing convergence.

However, RM fundamentally alters the waveforming architecture, breaking backward compatibility with existing systems, and the loss of structured channel sparsity makes channel estimation significantly more complex. 
To address this gap, we propose \emph{random precoding} (RP), a novel framework that integrates a random linear transform before the waveform modulator of any existing system. RP statistically exploits all subchannels for diversity gain while leaving the waveform, frame structure, and channel estimation entirely unchanged. Taking MIMO-OFDM as an example, the equivalent channel retains its block-diagonal structure in the frequency domain, which a low-complexity CD-OAMP detector exploits to achieve near-MAP-optimal performance.

The main contributions of this paper are as follows:
\begin{itemize}
    
    \item A universal RP framework that ensures the equivalent channel belongs to the universality class, achieving diversity gain and replica MAP-optimal detection while preserving waveform compatibility.
    \item Low-complexity RP constructions, including FT-RP and IBSFT-RP, with a compressed variant enabling flexible spectral efficiency.
    \item A CD-OAMP/MAMP detector for RP-MIMO systems, exploiting channel structure for reduced complexity.
    \item A state evolution analysis showing that CD-OAMP/MAMP achieves replica MAP-optimal BER under a unique fixed point.

\end{itemize}


\section{Preliminary}\label{sec:pre}

\subsection{System Model}

Consider a MIMO system with $N_{\rm tx}$ transmit and $N_{\rm rx}$ receive antennas and $K$ subcarriers. Let $N_T=KN_{\rm tx}$ and $N_R=KN_{\rm rx}$. After cyclic prefix (CP) removal, the received signal is
\begin{equation}
    \bm y = \bm H_{\rm MIMO}\bm x + \bm w,
\end{equation}
where $\bm x\in\mathbb{C}^{N_T\times 1}$ is the transmitted signal, $\bm{w} \sim \mathcal{CN}(\bm{0}, \sigma^2 \mathbf{I}_{N_R})$ is AWGN, and
\begin{equation}
    \bm H_{\rm MIMO} = \begin{bmatrix}
        \bm H_{1,1} & \cdots & \bm H_{1,N_{\rm tx}}\\
        \vdots & \ddots & \vdots\\
        \bm H_{N_{\rm rx},1} & \cdots & \bm H_{N_{\rm rx},N_{\rm tx}}
    \end{bmatrix}
\end{equation}
with $\bm H_{n_r,n_t}\in\mathbb{C}^{K\times K}$ being the channel matrix from transmit antenna $n_t$ to receive antenna $n_r$.

For a waveform with modulation matrix $\bm M\in\mathbb{C}^{K\times K}$, the symbol vector $\bm s\in\mathbb{C}^{N_T\times 1}$ with i.i.d. entries from constellation $\mathcal{Q}$ is modulated as $\bm x=(\bm I_{N_{\rm tx}}\otimes\bm M)\bm s$. Demodulation $\tilde{\bm y}=(\bm I_{N_{\rm rx}}\otimes\bm M^{\rm H})\bm y$ yields the equivalent channel model
\begin{subequations}\label{eq:A_equiv}
\begin{align}
    \tilde{\bm y} &= \bm A_{\rm equiv}\bm s + \tilde{\bm w},\\
    \bm A_{\rm equiv} &= (\bm I_{N_{\rm rx}}\otimes\bm M^{\rm H})\bm H_{\rm MIMO}(\bm I_{N_{\rm tx}}\otimes\bm M).
\end{align}
\end{subequations}
Representative modulation matrices include $\bm{M} = \bm{F}_K^{\mathrm{H}}$ for OFDM, $\bm{M} = \bm{I} \otimes \bm{F}^{\mathrm{H}}$ for OTFS, and $\bm{M} = \bm{\Lambda}_{c_1}^{\mathrm{H}} \bm{F}_K^{\mathrm{H}} \bm{\Lambda}_{c_2}^{\mathrm{H}}$ for AFDM. For OFDM in quasi-static channels, $\bm A_{\rm equiv}$ is block-diagonal in the frequency domain with $K$ blocks $\{\bm A_k\in\mathbb{C}^{N_{\rm rx}\times N_{\rm tx}}\}_{k=1}^K$. For OTFS and AFDM, $\bm A_{\rm equiv}$ is sparse but not block-diagonal, which prevents complexity reduction via block-wise LMMSE.

\subsection{Universality Class}

\begin{definition}[Universality Class \cite{dudeja2024spectral}]\label{Def:Universality}
A matrix ensemble $\mathscr{U}$ is a \emph{universality class} if any $\bm{U}\in\mathscr{U}$ can be expressed as $\bm{U}=\bm{J}\bm{D}$, where: (1) $\bm{D}=\diag\{e^{i\theta_1},\ldots,e^{i\theta_N}\}$ with $\theta_{1:N}\overset{\rm i.i.d.}{\sim}\mathsf{Unif}[0,2\pi)$; (2) $\bm{J}$ is deterministic and spectrally convergent with $\|\bm{J}\|_2\lesssim 1$; (3) for any fixed $k\in\mathbb{N}^*$ and $\epsilon>0$,
\begin{equation}
    \Big\|(\bm{J}^{\rm H}\bm{J})^k - \tfrac{{\rm tr}[(\bm{J}^{\rm H}\bm{J})^k]}{N}\bm{I}_N\Big\|_{\max} \lesssim N^{-1/2+\epsilon}.
\end{equation}
\end{definition}

The universality class ensures that the dynamics of AMP-type iterative algorithms can be accurately tracked by state evolution (SE), which is the key to proving replica MAP optimality. We extend the original real-field definition of \cite{dudeja2024spectral} to the complex field and assume the key SE properties carry over.

\section{Random Precoding Framework}\label{sec:RP-MIMO-OFDM}

\subsection{Random Precoding}

\begin{definition}[RP]\label{Def:RP}
A random precoding (RP) is a random linear transform $\tilde{\bm{s}}=\bm{\Xi}\bm{s}$, where $\bm{\Xi}\in\mathbb{C}^{N\times M}$ satisfies: (1) $\bm\Xi$ is independent of signal $\bm s$ and channel $\bm A_{\rm equiv}$; (2) $\bm A_{\rm equiv}\bm\Xi$ belongs to the universality class $\mathscr{U}$.
\end{definition}

RP can be a random unitary transform ($N=M$), a random compression transform ($N<M$), or a random spreading transform ($N>M$). The randomly precoded signal $\tilde{\bm s}=\bm\Xi\bm s$ exhibits an asymptotic Gaussian distribution while preserving signal energy. Since $\bm A_{\rm equiv}\bm\Xi\in\mathscr{U}$, every symbol $s_i$ statistically experiences all subchannels, achieving full diversity gain. The precoded signal can be embedded directly as the data payload in any standard frame structure, ensuring backward compatibility.

\textbf{Practical Implementations.}  RP can be realized by a variety of constructions with different complexity-performance trade-offs. We focus on three representative ones below.

\subsubsection{Haar-RP}
A Haar-distributed unitary matrix provides the highest randomness, generated via QR decomposition of a complex Gaussian matrix. However, its complexity of $\mathcal{O}(N^2)$ and memory cost make it impractical for large systems; it serves as a theoretical benchmark.

\subsubsection{FT-RP}
The randomly permuted fast transform achieves near-Haar performance with only $\mathcal{O}(N\log N)$ complexity:
\begin{equation}
    \bm\Xi = \bm\Pi\bm T_N,
\end{equation}
where $\bm\Pi\in\{0,1\}^{N\times N}$ is a random permutation matrix and $\bm T_N$ is a fast unitary transform such as FFT or Walsh-Hadamard transform (WHT). Memory requirements are only $\mathcal{O}(N)$.

\subsubsection{IBSFT-RP}
The interleaved block-sparse fast transform (IBSFT) \cite{IBS} provides further reduced complexity. It decomposes the vector into $L$ blocks of size $N_s\times M_s$ (with $N=LN_s$, $M=LM_s$) and applies localized fast transforms with interleaving:
\begin{equation}\label{eq:IBSFT}
    \bm\Xi = \bm\Pi\begin{bmatrix}
        \bm\Pi_1\bm T_{M_s} & \cdots & 0\\
        \vdots & \ddots & \vdots\\
        0 & \cdots & \bm\Pi_L\bm T_{M_s}
    \end{bmatrix},
\end{equation}
where $\bm\Pi_l$ are inner permutation matrices and $\bm\Pi$ is an outer permutation. When $N_s=M_s$, IBSFT-RP is a random unitary transform with complexity of $\mathcal{O}(N\log N_s)$, reduced from $\mathcal{O}(N\log N)$ by a factor $\frac{\log N_s}{\log N}$; it degenerates to FT-RP when $L=1$. When $N_s<M_s$, the incomplete inner permutation matrices realize random compression with ratio $\delta=N_s/M_s$, enabling flexible spectral efficiency control.

\subsection{RP-MIMO System}

\begin{figure}[t]
    \centering
    \includegraphics[width=\linewidth]{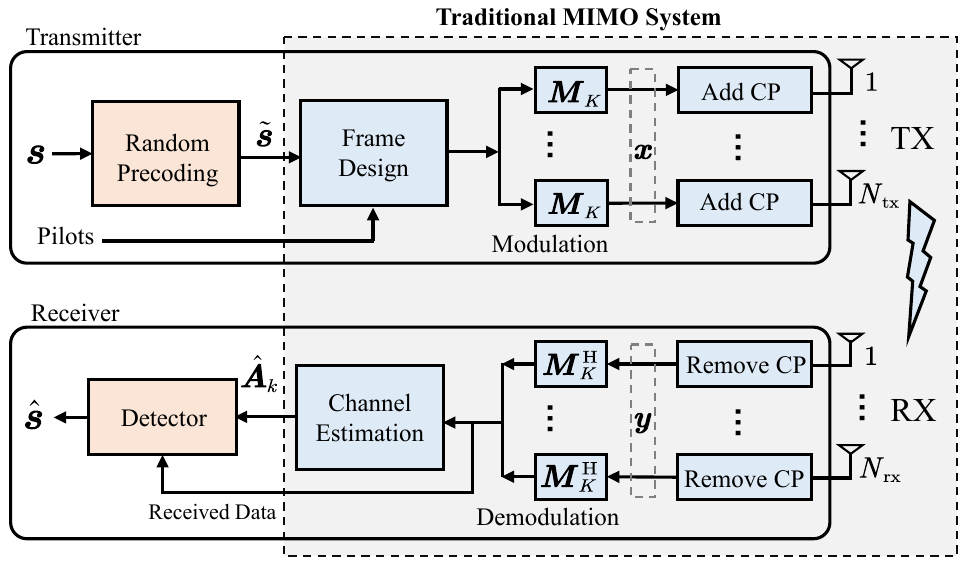}
    \caption{Framework of an RP-MIMO system. RP is inserted before the waveform modulator, leaving the waveform and receiver unchanged.}
    \label{fig:MIMO}
\end{figure}

As shown in Fig.~\ref{fig:MIMO}, the RP-MIMO system applies RP before waveforming: $\tilde{\bm s}=\bm\Xi\bm s$, $\bm\Xi\in\mathbb{C}^{N_T\times M_T}$. Substituting into \eqref{eq:A_equiv}, the signal detection problem for RP-MIMO becomes
\begin{subequations}\label{eq:RP-MIMO model}
\begin{align}
    \Gamma:&\quad \tilde{\bm y} = \bm A_{\rm equiv}\bm\Xi\bm s + \tilde{\bm w},\\
    \Phi:&\quad s_i\sim P_s,
\end{align}
\end{subequations}
where $\Gamma$ captures the linear channel model and $\Phi$ the discrete symbol prior. Since RP operates entirely before the waveform modulator, the waveform, frame structure, pilot design, and receiver demodulation remain unchanged. Channel estimation uses the same pilot-based method as the host waveform (e.g., LS or LMMSE in the frequency domain for OFDM), which is a fundamental advantage over RM.

Compared with RM \cite{Liu2025RandomMultiplexing}, RP offers three key advantages:
\begin{itemize}
    \item \textbf{Waveform compatibility}: RP only adds a precoding block before modulation, without altering the existing transmitter architecture.
    \item \textbf{Channel estimation}: RP inherits the host waveform's pilot design and channel estimator unchanged, avoiding the need for new estimation schemes.
    \item \textbf{Detection complexity}: RP preserves the sparse structure of $\bm A_{\rm equiv}$ (e.g., block-diagonal in MIMO-OFDM), enabling further complexity reduction in signal detection.
\end{itemize}

For MIMO-OFDM in quasi-static channels, the equivalent RP-MIMO-OFDM model is
\begin{equation}\label{eq:MIMO2}
    \tilde{\bm y} = \bar{\bm\Pi}^{\rm T}{\rm diag}(\bm A_1,\ldots,\bm A_K)\bar{\bm\Pi}\bm\Xi\bm s + \tilde{\bm w},
\end{equation}
where $\bm A_k\in\mathbb{C}^{N_{\rm rx}\times N_{\rm tx}}$ is the per-subcarrier channel matrix and $\bar{\bm\Pi}$ is an antenna-subcarrier permutation. This block-diagonal structure is central to CD-OAMP's complexity reduction.

\section{CD-OAMP/MAMP Detection}\label{sec:OAMP}

In this section, we study the CD-OAMP/MAMP detector for RP-MIMO systems. CD-OAMP/MAMP is composed of a linear estimator (LE) in the frequency domain and a nonlinear estimator (NLE) in the random precoding domain, connected by random transform and its inverse. CD-OAMP is particularly effective for quasi-static channels where $\bm A_{\rm equiv}$ is block-diagonal, while CD-MAMP handles more general scenarios by replacing the LMMSE with a memory-based estimator.

\subsection{CD-OAMP for Unitary RP}

\begin{figure}[t]
    \centering
    \includegraphics[width=0.9\linewidth]{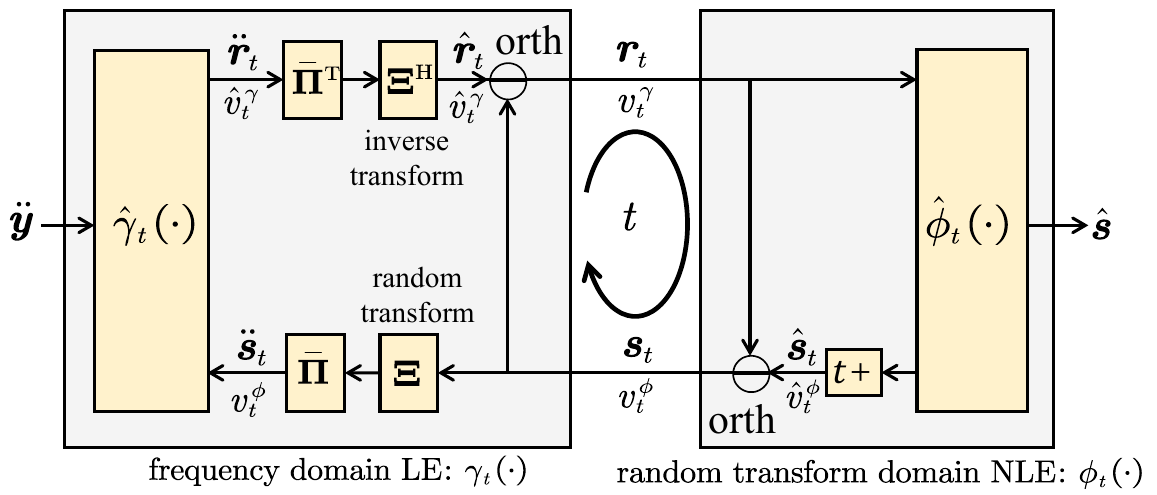}
    \caption{CD-OAMP detector for unitary RP systems: LE in the frequency domain and MMSE NLE in the random precoding domain. Random transform and its inverse transfer information between the two domains.}
    \label{fig:CD-OAMP}
\end{figure}

The detection problem \eqref{eq:RP-MIMO model} is solved by CD-OAMP through two local estimators targeting $\Gamma$ and $\Phi$ respectively, as shown in Fig.~\ref{fig:CD-OAMP}. Symbol vector $\bm s$ lives in the random precoding domain, while $\ddot{\bm s}=\bar{\bm\Pi}\bm\Xi\bm s$ lies in the frequency domain. Denote the error terms at the output of LE and NLE after orthogonalization as $\bm g_t=\bm r_t-\bm s$ and $\bm f_t=\bm s_t-\bm s$, with per-symbol MSEs
\begin{subequations}
\begin{align}
    v_t^\gamma &= \tfrac{1}{N_T}{\rm E}\{\|\bm g_t\|^2\}, \quad v_t^\phi = \tfrac{1}{N_T}{\rm E}\{\|\bm f_t\|^2\}.
\end{align}
\end{subequations}

\subsubsection{LE}
The LE performs LMMSE estimation of $\ddot{\bm s}$ in the frequency domain. After orthogonalization and random transform, the message $\ddot{\bm s}_t$ satisfies $\ddot{\bm s}\sim\mathcal{CN}(\ddot{\bm s}_t, v_t^\phi\bm I_{N_T})$. Exploiting the block-diagonal structure of \eqref{eq:MIMO2}, the large-scale LMMSE decomposes into $K$ independent per-subcarrier LMMSEs of size $N_{\rm rx}\times N_{\rm tx}$. The weight matrix $\bm W_t = \diag (\bm W_{1,t},\cdots, \bm W_{k,t})$ is also block-diagonal:
\begin{subequations}\label{eq:LMMSE}
\begin{align}
    \bm W_{k,t} &= v_t^\phi\bm A_k^{\rm H}(\sigma^2\bm I + v_t^\phi\bm A_k\bm A_k^{\rm H})^{-1},\label{eq:W_kt}\\
    \ddot{\bm r}_{k,t} &= \ddot{\bm s}_{k,t} + \bm W_{k,t}(\ddot{\bm y}_k - \bm A_k\ddot{\bm s}_{k,t}),\\
    \hat v_t^\gamma &= \frac{v_t^\phi}{K}\sum_{k=1}^K\Big(1-\frac{{\rm tr}\{\bm W_{k,t}\bm A_k\}}{N_{\rm tx}}\Big).
\end{align}
\end{subequations}
This reduces per-iteration complexity from $\mathcal{O}(N_R^3)$ to $\mathcal{O}(KN_{\rm rx}^3)=\mathcal{O}(N_TN_{\rm tx}^2)$.

\subsubsection{IRT and RT}
After LE, the inverse random transform (IRT) maps $\ddot{\bm r}_t$ from the frequency domain to the random precoding domain:
\begin{equation}
    {\rm IRT:}\quad\hat{\bm r}_{t} = \bm\Xi^{\rm H}\bar{\bm\Pi}^{\rm T}\ddot{\bm r}_{t}.
\end{equation}
The forward random transform (RT) before LE converts $\bm s_t$ back to the frequency domain:
\begin{equation}
    {\rm RT:}\quad\ddot{\bm s}_{t} = \bar{\bm\Pi}\bm\Xi{\bm s}_{t}.
\end{equation}
For unitary $\bm\Xi$, both transforms preserve error variances. The random transform also promotes i.i.d. Gaussianity of estimation errors \cite{liu2023oamp}, so the NLE input satisfies $\bm r_t\sim\mathcal{CN}(\bm s,v_t^\gamma\bm I)$.

\subsubsection{NLE}
The NLE performs symbol-by-symbol MMSE demodulation in the random precoding domain over constellation $\mathcal{Q}=\{a_1,\ldots,a_Q\}$:
\begin{subequations}
\begin{align}
    \hat{\bm s}_{t+1} &= \hat\phi_t(\bm r_t) = {\rm E}\{\bm s\,|\,\bm r_t,\,\bm s\in\mathcal{Q}\},\\
    \hat v_{t+1}^\phi &= {\rm Var}\{\bm s\,|\,\bm r_t,\,\bm s\in\mathcal{Q}\},
\end{align}
\end{subequations}
with element-wise soft-decision computation of complexity $\mathcal{O}(N_T)$.

\subsubsection{Orthogonalization}
Standard OAMP orthogonalization \cite{ma2017orthogonal} eliminates input-output error correlation:
\begin{subequations}\label{eq:orth}
\begin{align}
    v_t^\gamma &= [(\hat v_t^\gamma)^{-1}-(v_t^\phi)^{-1}]^{-1},\\
    \bm r_t &= (v_t^\gamma)^{-1}[(\hat v_t^\gamma)^{-1}\hat{\bm r}_t-(v_t^\phi)^{-1}\bm s_t],\\
    v_{t+1}^\phi &= [(\hat v_{t+1}^\phi)^{-1}-(v_t^\gamma)^{-1}]^{-1},\\
    \bm s_{t+1} &= (v_{t+1}^\phi)^{-1}[(\hat v_{t+1}^\phi)^{-1}\hat{\bm s}_{t+1}-(v_t^\gamma)^{-1}\bm r_t].
\end{align}
\end{subequations}
The RP and orthogonalization operations together ensure the following orthogonality conditions for all $t\ge 1$ as $N_T\to\infty$: $\frac{1}{N_T}\bm s^{\rm H}\bm g_t\to 0$, $\frac{1}{N_T}\bm f_t^{\rm H}\bm g_t\to 0$, and $\frac{1}{N_T}\bm g_t^{\rm H}\bm f_{t+1}\to 0$. These are the key conditions for state evolution to hold. 

\subsubsection{Relation to Original OAMP}
In fact, performing LMMSE for $\ddot{\bm s}$ in the frequency domain combined with IRT/RT is equivalent to performing LMMSE for $\bm s$ directly via the dense equivalent channel $\bm C=\bm A\bar{\bm\Pi}\bm\Xi$:
\begin{equation}
    \bar{\bm\Xi}^{\rm H}\hat\gamma_t(\bar{\bm\Xi}\bm s_t) = \bm s_t + v_t^\phi\bm C^{\rm H}(\sigma^2\bm I+v_t^\phi\bm{CC}^{\rm H})^{-1}(\ddot{\bm y}-\bm C\bm s_t),
\end{equation}
where $\bar{\bm\Xi}=\bar{\bm\Pi}\bm\Xi$. The key advantages of CD-OAMP over this direct form are: (i) the block-diagonal structure of $\bm A$ avoids dense large-scale matrix inversion, drastically reducing complexity; (ii) the random transform enhances i.i.d. Gaussianity of estimation errors compared with conventional MIMO-OFDM without RP.

\subsubsection{State Evolution and Replica MAP Optimality}

\begin{lemma}[State Evolution]\label{lemma:SE}
Suppose $\bm A_{\rm equiv}\bm\Xi\in\mathscr{U}$ and $N_T\to\infty$. The SE of CD-OAMP holds almost surely:
\begin{subequations}
\begin{align}
    \hat v_t^\gamma &\overset{\rm a.s.}{=} \frac{v_t^\phi}{K}\sum_{k=1}^K\Big(1-\frac{{\rm tr}\{\bm W_{k,t}\bm A_k\}}{N_{\rm tx}}\Big),\\
    \hat v_{t+1}^\phi &\overset{\rm a.s.}{=} {\rm E}_{S,Z}\big\{|\hat\phi_t(S+\sqrt{v_t^\gamma}Z)-S|^2\big\},
\end{align}
\end{subequations}
where $S\sim P_s$ and $Z\sim\mathcal{CN}(0,1)$ are independent.
\end{lemma}

The SE recursion provides exact performance prediction without Monte Carlo simulation. RP ensures $\bm A_{\rm equiv}\bm\Xi\in\mathscr{U}$, the key condition for SE to hold.

\begin{lemma}[Replica MAP Optimality]\label{lemma:MAP}
Under the assumptions of Lemma~\ref{lemma:SE}, if the SE has a unique fixed point, CD-OAMP achieves the replica MAP-optimal BER \cite{ma2017orthogonal}.
\end{lemma}

\subsection{CD-MAMP for Unitary RP}

In scenarios where the equivalent channel is not block-diagonal (e.g., MIMO-OTFS and MIMO-AFDM), or in large-scale MIMO-OFDM with large $N_{\rm tx}$ where the per-subcarrier LMMSE cost $\mathcal{O}(N_T N_{\rm tx}^2)$ becomes prohibitive, CD-MAMP \cite{liu2022memory} replaces LMMSE with a memory linear estimator (MLE).

\subsubsection{Memory LE}
Starting with $t=1$ and $\tilde{\bm S}_1=\bm 0$, the MLE depends on all previous NLE outputs $\tilde{\bm S}_t=[\tilde{\bm s}_1,\ldots,\tilde{\bm s}_t]$:
\begin{equation}
    \ddot{\bm r}_t = \gamma_t(\tilde{\bm S}_t)=\frac{1}{\epsilon_t^\gamma}\big(\hat\gamma_t(\tilde{\bm S}_t)-\tilde{\bm S}_t\bm p_t\big),
\end{equation}
where $\hat\gamma_t(\tilde{\bm S}_t)=\bm A_{\rm equiv}^{\rm H}\tilde\gamma_t(\tilde{\bm S}_t)$ with recursion
\begin{equation}
    \tilde\gamma_t(\tilde{\bm S}_t)=\bm B_t\tilde\gamma_{t-1}(\tilde{\bm S}_{t-1}) + \xi_t(\tilde{\bm y}-\bm A_{\rm equiv}\tilde{\bm s}_t),
\end{equation}
$\tilde\gamma_0=\bm 0$, and $\bm B_t=\theta_t(\lambda^\dagger\bm I-\bm A_{\rm equiv}\bm A_{\rm equiv}^{\rm H})$.

\subsubsection{Memory NLE}
A damping vector $\bm\zeta_t$ is applied to combine past estimates:
\begin{equation}
    \ddot{\bm s}_t = [\ddot{\bm s}_1,\ldots,\ddot{\bm s}_{t-1},\tilde{\bm s}_t]\bm\zeta_t,
\end{equation}
where parameters are chosen to ensure state evolution and replica MAP optimality. For sparse $\bm A_{\rm equiv}$ with sparsity rate $\rho$, the MLE requires $\mathcal{O}(\rho N_T^2)$ complexity per iteration.

\vspace{-8mm}\subsection{CD-OAMP for Compressed IBSFT-RP}

\begin{figure}[t]
    \centering
    \includegraphics[width=\linewidth]{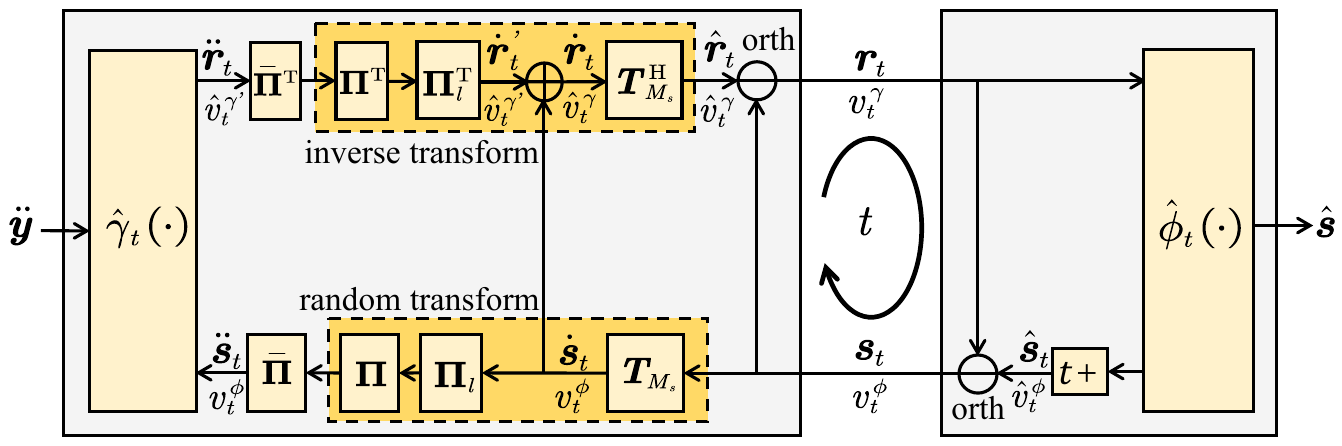}
    \caption{CD-OAMP detector for compressed IBSFT-RP systems.}
    \label{fig:CD-OAMP-cmp}
\end{figure}



For compressed RP ($N<M$, $\delta=N/M<1$), the inverse transform is non-invertible since the entries discarded by the incomplete inner permutation matrices $\bm\Pi_l$ in \eqref{eq:IBSFT} are lost. As shown in Fig.~\ref{fig:CD-OAMP-cmp}, the lost entries are padded from the current NLE estimate $\dot{\bm s}_t$: $\dot{\bm r}_t=\dot{\bm r}'_t+\bm p_t$, where $\dot{\bm r}'_t$ is the partial IRT output and $\bm p_t$ copies the unobserved entries from $\dot{\bm s}_t$. Thus, $\hat v_t^\gamma=\delta\hat v_t^{\gamma'}+(1-\delta)v_t^\phi$, with $\hat v_t^{\gamma'}$ denoting the LE output variance. The forward random transform involves no padding and hence preserves the variance. Other steps are identical to the unitary RP case and extend directly to CD-MAMP.

\subsection{Complexity Comparison}

Table~\ref{tab:complexity} compares per-iteration complexity. FT-RP with CD-OAMP achieves $\mathcal{O}(N_T\log N_T+N_TN_{\rm tx}^2)$, substantially lower than CD-OAMP for OTFS/AFDM with $\mathcal{O}(N_T^3)$. IBSFT-RP further reduces the transform cost to $\mathcal{O}(N_T\log N_s)$. For massive MIMO with large $N_{\rm tx}$, CD-MAMP's $\mathcal{O}(\rho N_T^2)$ MLE cost scales more favorably than CD-OAMP's $\mathcal{O}(N_TN_{\rm tx}^2)$ LMMSE.

\renewcommand{\arraystretch}{1.3}
\begin{table}[t]
    \centering
    \caption{Per-Iteration Complexity of Detectors in MIMO Systems}
    \begin{tabular}{l|l}
    \hline
    Scheme & Complexity\\
    \hline
    OFDM: LMMSE+MMSE & $\mathcal{O}(N_TN_{\rm tx}^2+N_T)$\\
    OTFS/AFDM: CD/DD-OAMP \cite{9536449,10043628} & $\mathcal{O}(N_T^3+N_T)$\\
    IFDM: CD-MAMP \cite{10522098} & $\mathcal{O}(\rho N_T^2+N_T\log N_T)$\\
    \hline
    FT-RP-OFDM: CD-OAMP (proposed) & $\mathcal{O}(N_T\log N_T+N_TN_{\rm tx}^2)$\\
    IBSFT-RP-OFDM: CD-OAMP (proposed) & $\mathcal{O}(N_T\log N_s+N_TN_{\rm tx}^2)$\\
    FT-RP-OFDM: CD-MAMP (proposed) & $\mathcal{O}(N_T\log N_T+\rho N_T^2)$\\
    IBSFT-RP-OFDM: CD-MAMP (proposed) & $\mathcal{O}(N_T\log N_s+\rho N_T^2)$\\
    \hline
    \end{tabular}
    \label{tab:complexity}
\end{table}

\section{Numerical Results}\label{sec:num}

Simulations compare RP with original OFDM, OTFS, AFDM, and IFDM. A $2\times2$ MIMO system with $K=400$ subcarriers at $\Delta f=15\,\text{kHz}$ is used. A TDL channel with 300\,ns delay spread generates quasi-static multipath fading with severe frequency-selective behavior. Data and RP matrices are regenerated independently for each trial. Both CD-OAMP and CD-MAMP use $\mathcal{T}=10$ iterations.
SNR is defined per receive antenna, i.e., $\mathrm{SNR} = \frac{\mathbb{E}[|\bm{H}\bm{x}|^2]}{N_{\mathrm{rx}} \sigma^2}$.

\begin{figure}[t]
    \centering
    \includegraphics[width=0.9\linewidth]{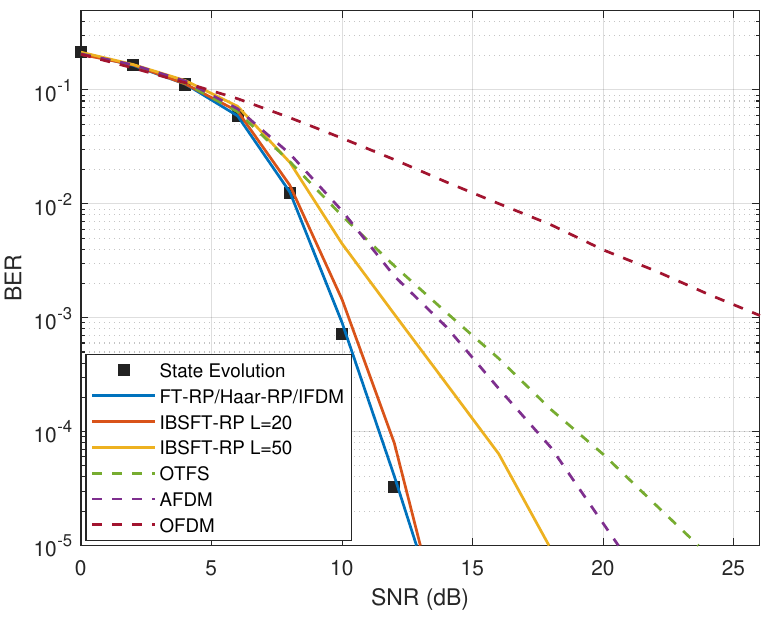}
    \caption{BER comparison over QPSK in $2\times2$ MIMO, $K=400$. CD-OAMP is used for RP-OFDM, OTFS, AFDM; CD-MAMP for IFDM.}
    \label{fig:RP_plot_QPSK}
\end{figure}

\begin{figure}[t]
    \centering
    \includegraphics[width=0.9\linewidth]{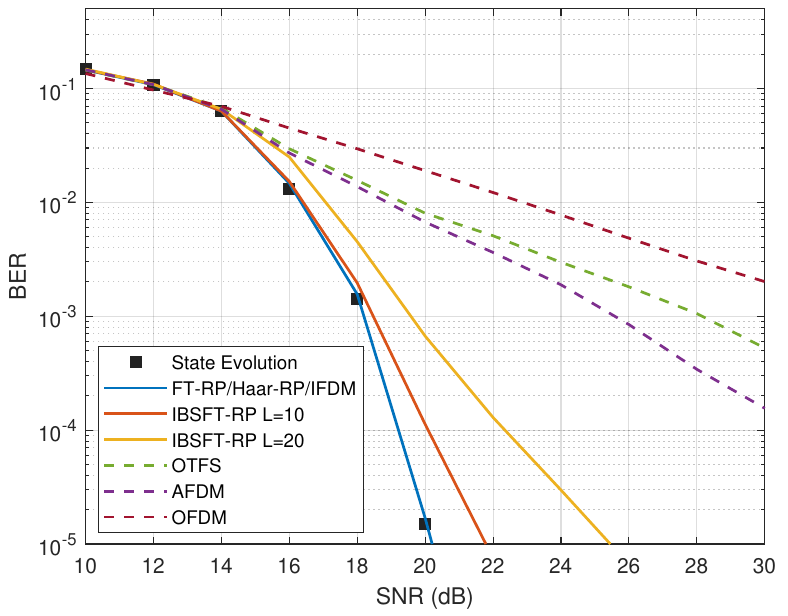}
    \caption{BER comparison over 16QAM in $2\times2$ MIMO, $K=400$.}
    \label{fig:RP_plot_16QAM}
\end{figure}


Figs.~\ref{fig:RP_plot_QPSK} and \ref{fig:RP_plot_16QAM} show BER results under QPSK and 16QAM. All RP configurations achieve significant diversity gain over OFDM, OTFS, and AFDM, as each symbol statistically experiences all subchannels. FT-RP and Haar-RP achieve near-identical BER to IFDM/RM, confirming that RP inherits RM's diversity gain without altering the waveform, since both ensure $\bm A_{\rm equiv}\in\mathscr{U}$ with unchanged eigenvalue distribution. IBSFT-RP incurs a moderate performance loss that increases with the number of blocks $L$, due to reduced transform dimension and weaker inter-symbol spreading. Nevertheless, IBSFT-RP with $L=20$ still outperforms existing waveforms. The SE-predicted BER closely matches simulation results, validating Lemma~\ref{lemma:SE}.

\begin{figure}[t]
    \centering
    \includegraphics[width=0.9\linewidth]{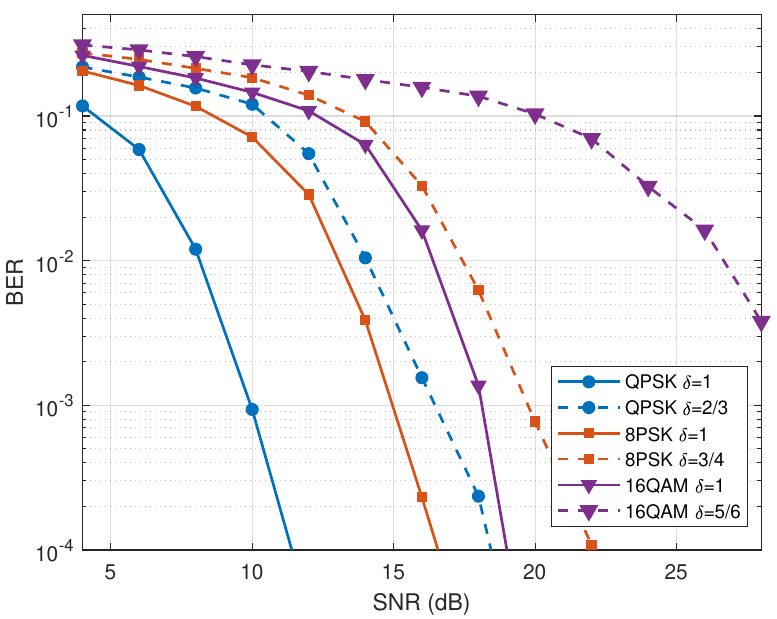}
    \caption{Compressed FT-RP-MIMO-OFDM with QPSK, $N_{\rm tx}=N_{\rm rx}=2$, $K=600$, $\delta=KN_{\rm tx}/M_T$.}
    \label{fig:compression}
\end{figure}

Fig.~\ref{fig:compression} evaluates compressed RP. With $\delta=2/3$, QPSK achieves the same spectral efficiency as 8PSK; with $\delta=3/4$, 8PSK matches 16QAM. The BER degradation is moderate and controlled by adjusting $\delta$, providing a flexible mechanism to trade performance for spectral efficiency without changing the modulation order.

\begin{figure}[t]
    \centering
    \includegraphics[width=0.9\linewidth]{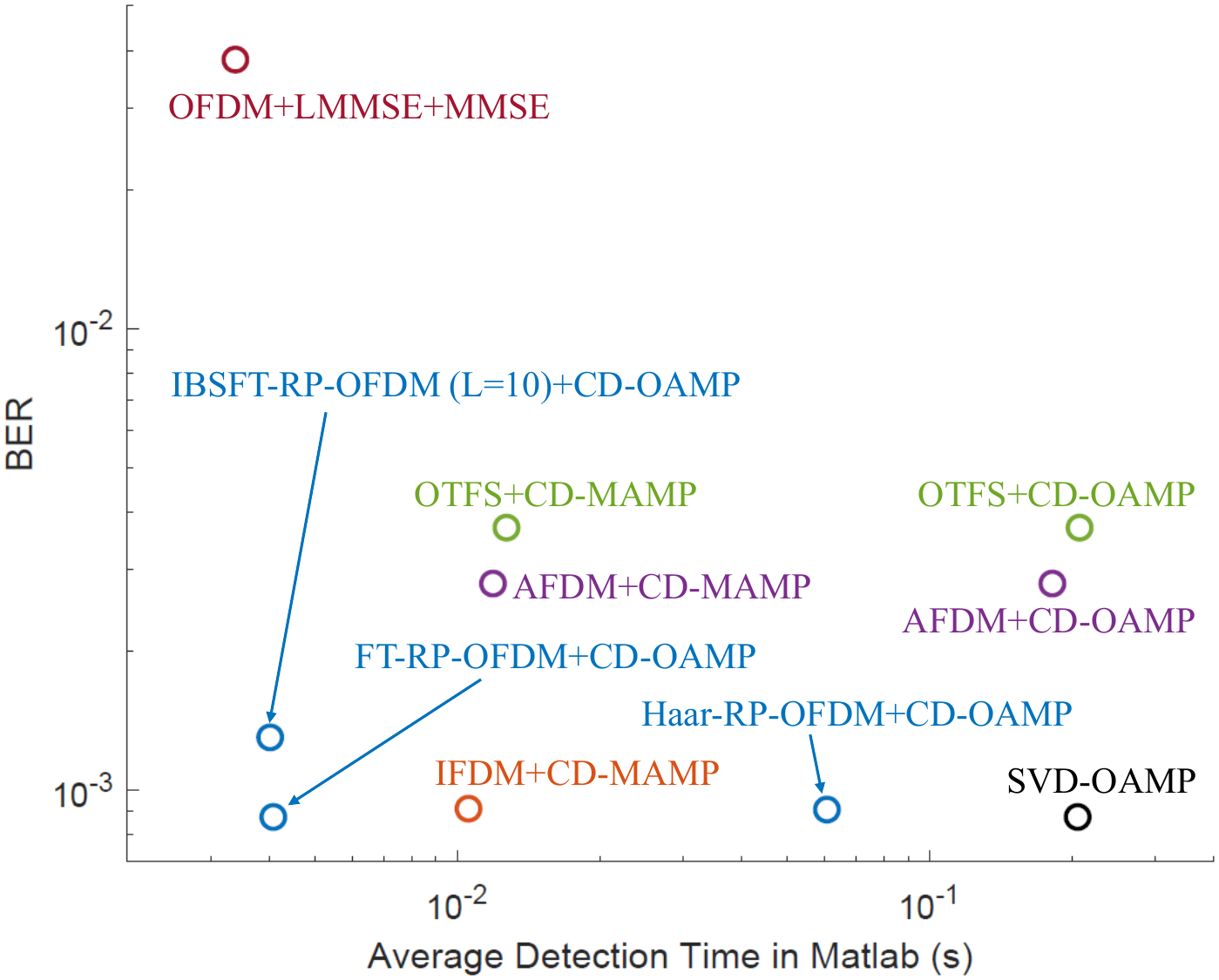}
    \caption{BER vs. detection time (Matlab) in $2\times2$ MIMO, QPSK, $K=400$, SNR$=10\,$dB.}
    \label{fig:timer}
\end{figure}

Fig.~\ref{fig:timer} plots BER against average detection time. FT-RP with CD-OAMP achieves the best BER with very low detection time, far outperforming OTFS/AFDM with CD-OAMP that require large-scale matrix inversion. IBSFT-RP reduces detection time further at a slight BER cost, beneficial for hardware-constrained deployments. For massive MIMO, CD-MAMP provides a scalable alternative due to its quadratic complexity scaling.\vspace{-1mm}

\section{Conclusion}\label{sec:conc}\vspace{-1mm}

This paper has proposed a universal random precoding (RP) framework that statistically exploits all subchannels through random linear transforms, achieving significant diversity gains while maintaining full backward compatibility with existing waveforms. A CD-OAMP/MAMP detector is designed for RP-MIMO systems, exploiting the block-diagonal frequency-domain structure to achieve near MAP-optimal performance validated by state evolution. Efficient FT-RP and IBSFT-RP constructions offer low complexity, and compressed RP enables flexible spectral efficiency. Numerical results confirm that RP outperforms existing waveforms with lower detection complexity, positioning it as a versatile solution for next-generation wireless communications.

\bibliographystyle{IEEEtran}
\bibliography{reference}

\end{document}